\documentclass[12pt,a4paper]{article}
\usepackage[T1]{fontenc}
\usepackage[preprint]{dfttrob}
\usepackage{latexuseful2e}
\usepackage{amsmath}
\usepackage{graphicx}

\dfttnum{FISIST/10-99/CENTRA\\July 12, 1999}

\begin{document}

\newcommand{\Jpsi}{J/$\psi$}

\title{\bf Absorption and percolation in the production \\of \Jpsi\ in
heavy ion collisions}
\author{J. Dias de Deus, R. Ugoccioni and A. Rodrigues\\[2mm]
 \small\it CENTRA and Departamento de F{\'\i}sica (I.S.T.),\\ 
 \it Av. Rovisco Pais, 1049-001 Lisboa, Portugal}
\maketitle

\begin{abstract}
We present a simple model with string absorption and percolation to
describe the \Jpsi\ suppression in heavy ion collisions.
The NA50 data are fairly well explained by the model.
\end{abstract}

It was recently argued that central distributions, in transverse
energy $E_T$ or multiplicity $n$, are dominated by fluctuations in the
number $\nu$ of elementary collisions, in the case of nucleus-nucleus
interactions and even hadron-hadron interactions at very high energy
\cite{components}.

In that framework, if the total number of events with $\nu$ elementary
collisions is $N(\nu)$ and $N_C(\nu)$ is the number of events with the
rare event $C$ occurring, then \cite{rare}
\begin{equation}
  N_C(\nu) = \alpha_C \nu N(\nu) ,		\label{eq:rare}
\end{equation}
where $\alpha_C$ is the probability of event $C$ occurring in an
elementary collision. An event is rare if, in good approximation, it
only occurs once.

Relation \eqref{eq:rare} is correct as far as event $C$ is free
from absorption \cite{nuclear_absorption}, or from quark-gluon plasma
formation \cite{QGP}. This is directly seen in the observed linear
relation between $E_T$ distributions associated to Drell-Yan
productions, $N_{\text{D.Y.}}(E_T)$, and to minimum bias events, $N(E_T)$
\cite{rare,NA50:Jpsi}.

However, in the case of \Jpsi\ production, the formed \Jpsi\ may be
destroyed by subsequent interactions, as it moves through the
interacting medium \cite{Jpsi:absorption}, or it may happen that it is
not really formed, as a result of quark-gluon plasma Debye screening
preventing the $c\bar c$ binding.

In the standard multicollision model ---dual parton model, \cite{DPM:1},
for instance--- in every elementary collision two coloured strings are
originated. These strings constitute the medium that may absorb a
created \Jpsi\ (see also \cite{ccstrings}). 
They may also fuse (percolate) and form the
quark-gluon plasma, thus inhibiting \Jpsi\ creation.

We shall consider that when $\nu$ collisions occur they occupy an
interaction volume $V(\nu)$ characterised by a transverse area
$A(\nu)$ and a mean longitudinal length $L(\nu)$,
\begin{equation}
  V(\nu) \equiv A(\nu) L(\nu).		\label{eq:volume}
\end{equation}
As nuclear matter density is uniform, we expect the string density
$\rho_s$,
\begin{equation}
  \rho_s \equiv \frac{2\nu}{A(\nu) L(\nu)},		\label{eq:rho}
\end{equation}
where $2\nu$ is the number of formed strings, to be as well
uniform. Note that $\rho_s$ in eq.~\eqref{eq:rho} is, in general, independent
of the interacting nuclei, but may be energy dependent.

In order to take care of \Jpsi\ absorption we introduce the survival
probability $P_s(\nu)$ in the conventional form
\begin{equation}
  P_s(\nu) = \exp( -L(\nu) \rho_s \sigma )	,		\label{eq:Ps}
\end{equation}
$\sigma$ being the \Jpsi-string ($q\bar q$) absorption cross-section.

Next, we include the probability of quark-gluon plasma formation in a
very simple two-dimensional percolation model
\cite{perc:previous,Pajares:perc}.
In a $\nu$ elementary collision event there is a probability
$P_{np}(\nu)$ of percolation not occurring and a probability
$1-P_{np}(\nu)$ of percolation occurring.
As percolation means here quark-gluon plasma formation, we make the
strong assumption that the \Jpsi\ is formed only in events 
in which there is no percolation.
This means that in eq.~\eqref{eq:rare} we shall multiply $N_C(\nu)$ by
$P_{np}(\nu)$:
\begin{equation}
	N_C(\nu) \longrightarrow N_C(\nu) P_{np}(\nu)  .
\end{equation}
This probability $P_{np}(\nu)$ is in fact a function of the scaling
variable $\eta$, the dimensionless density of strings in the
transverse plane,
\begin{equation}
  \eta \equiv \pi r_s^2 \frac{2\nu}{A(\nu)},
\end{equation}
where $\pi r_s^2$ is the transverse area of a string. 
Because of eq.~\eqref{eq:rho}, we can also write
\begin{equation}
  \eta = \pi r_s^2 \rho_s L(\nu).					\label{eq:eta}
\end{equation}

\begin{figure}
\begin{center}
\mbox{\includegraphics[width=0.7\textwidth]{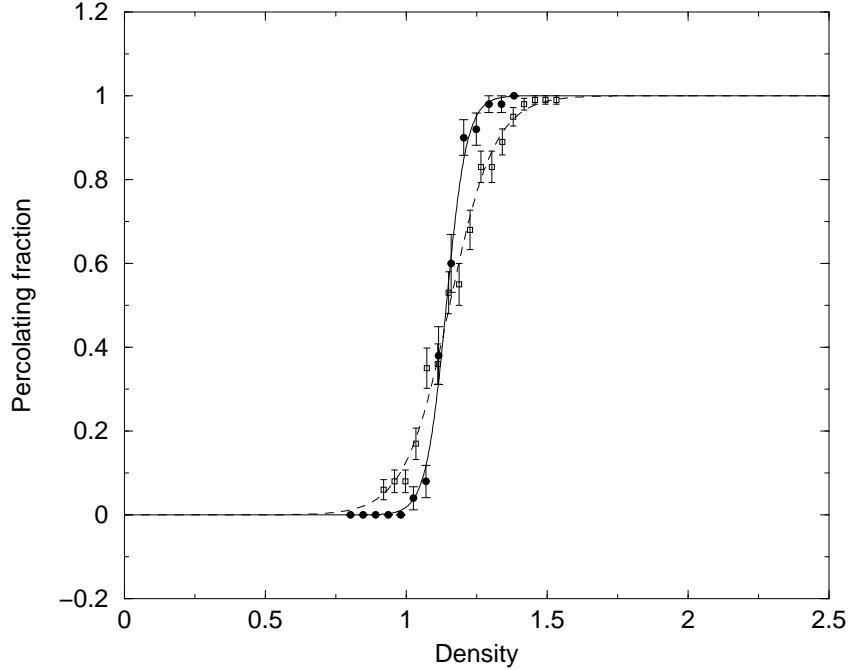}}
\end{center}
\caption[percolation probability]{Percolation probability as a
function of the transverse density $\eta$ for S-U
(open squares) and Pb-Pb (filled circles) collisions geometry as
obtained from Monte Carlo simulations \cite{perc:98}. 
The curves show fits with
$P_p(\eta) = 1 - P_{np}(\eta)$, see eq.~\eqref{eq:Pnp};
we obtain $a=0.07\pm 0.01$ for S-U (dashed line) and
$a=0.04\pm 0.01$ for Pb-Pb (solid line);
$\eta_c = 1.15\pm 0.02$}\label{fig:a}
\end{figure}

Recently, in \cite{perc:98}, the probability $P_{np}$ was studied
for S-U and Pb-Pb collisions at $\sqrt{s} = 19.4$ A\,GeV. It can be
written in the form
\begin{equation}
  P_{np}(\eta) = \frac{1}{\exp\left( \frac{\eta-\eta_c}{a} \right) +
  1}															\label{eq:Pnp}
\end{equation}
where $\eta_c$ is the critical density, $\eta_c \approx 1.15$ for a
uniform distribution of strings in the transverse plane, and $a$ is a
parameter which depends only on the geometry of the colliding nuclei
(see Figure~\ref{fig:a}).

We can now write our final formula for \Jpsi\ production, normalised to
Drell-Yan production, $R \equiv$ (\Jpsi)/(D.Y.), as
\begin{equation}
  R(\nu) = K P_s(\nu) P_{np}(\nu)		\label{eq:RL}
\end{equation}
where $K$ is a normalisation constant, which should be close to 50
from $pp$ collisions \cite{NA50:pp}. This corresponds to the small
$\nu$, small $E_T$ limit, with $P_s(\nu) \to 1$, $P_{np}(\nu) \to
1$. Validity of eq.~\eqref{eq:RL} requires \Jpsi\ and Drell-Yan
production being treated as rare events.

Now we note that, from eq.~\eqref{eq:Ps} and eq.~\eqref{eq:eta}, we
obtain that $R$ is in fact a function of the mean length $L$ only,
\begin{equation}
  R = K \exp( - L \rho_s \sigma )
		\left[\exp\left( 
			\frac{\pi r_s^2 \rho_s L-\eta_c}{a} \right) + 1\right]^{-1}
																			\label{eq:Rfull}
\end{equation}
For small values of L, away from percolation, the ratio $R$ in 
eq.~\eqref{eq:Rfull} is universal, i.e., does not depend on the
colliding nuclei, but it depends, in our approach, on the c.m.\
energy, through $\rho_s$. 

\begin{figure}
\begin{center}
\mbox{\includegraphics[width=0.91\textwidth]{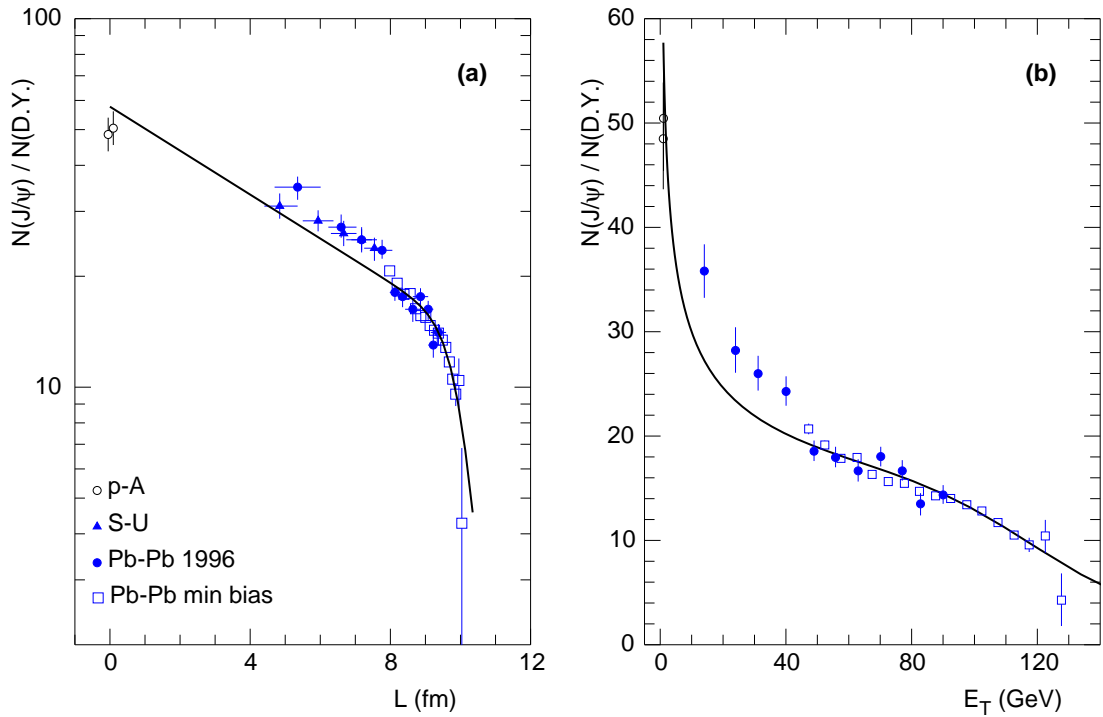}}
\end{center}
\caption[comparison with data]{The ratio of \Jpsi to Drell-Yan events
as predicted by eq.~\eqref{eq:RL}, compared to experimental data
as published by NA50 Collaboration 
\cite{NA50:Jpsi,NA50:QM99}.}\label{fig:Jpsidata}
\end{figure}

Results were presented by the NA50 Collaboration \cite{NA50:Jpsi} 
on S-U and Pb-Pb
collisions as functions of the average interaction distance $L$
computed in a geometrical model \cite{NA50:L,ccbarlength}.
In Figure~\ref{fig:Jpsidata}a we compare eq.~\eqref{eq:Rfull}
with the NA50 data. The string radius was fixed at the value $r_s=0.2$
fm, \cite{Pajares:perc}, which places $\eta$ for S-U somewhat below
the percolation threshold, and the parameter $a$ at the Pb-Pb value,
0.04.  
In Figure~\ref{fig:Jpsidata}a 
we have included the 1998 data on the ratio $R$ as a
function of $E_T$, and transformed to $L$ using the NA50 $E_T$ vs $L$
plot \cite{NA50:L} in the approximate form (valid for Pb-Pb at 19.4
AGeV),
\begin{equation}
  L \approx 2.07 \ln E_T ,		\label{eq:L-Et}
\end{equation}
with $E_T$ in units of GeV.

The curve shown in Figure~\ref{fig:Jpsidata}a corresponds to the
following values of the parameters in eq.~\eqref{eq:Rfull}: $\rho_s =
0.9$ fm$^{-3}$, $\sigma = 1.5$ mb and $K=58$.
The string density $\rho_s$ is, as expected \cite{DiasDeus:Jpsi}, 
much larger than nuclear density ($\rho \approx 0.17$ fm$^{-3}$) and
the absorption cross section agrees with what is measured in
\Jpsi-hadron collisions at intermediate energies.

In Figure~\ref{fig:Jpsidata}b we present the ratio $R$ as a function of
$E_T$ in comparison with data, by making use of eq.~\eqref{eq:L-Et}. 
The agreement is mostly qualitative.

The essential point is that at small $E_T$ the behaviour is typical of
absorption ---with positive curvature--- at higher values of $E_T$
the approach to the percolation transition produces a change in the
sign of the curvature.

\begin{figure}
\begin{center}
\mbox{\includegraphics[width=0.91\textwidth]{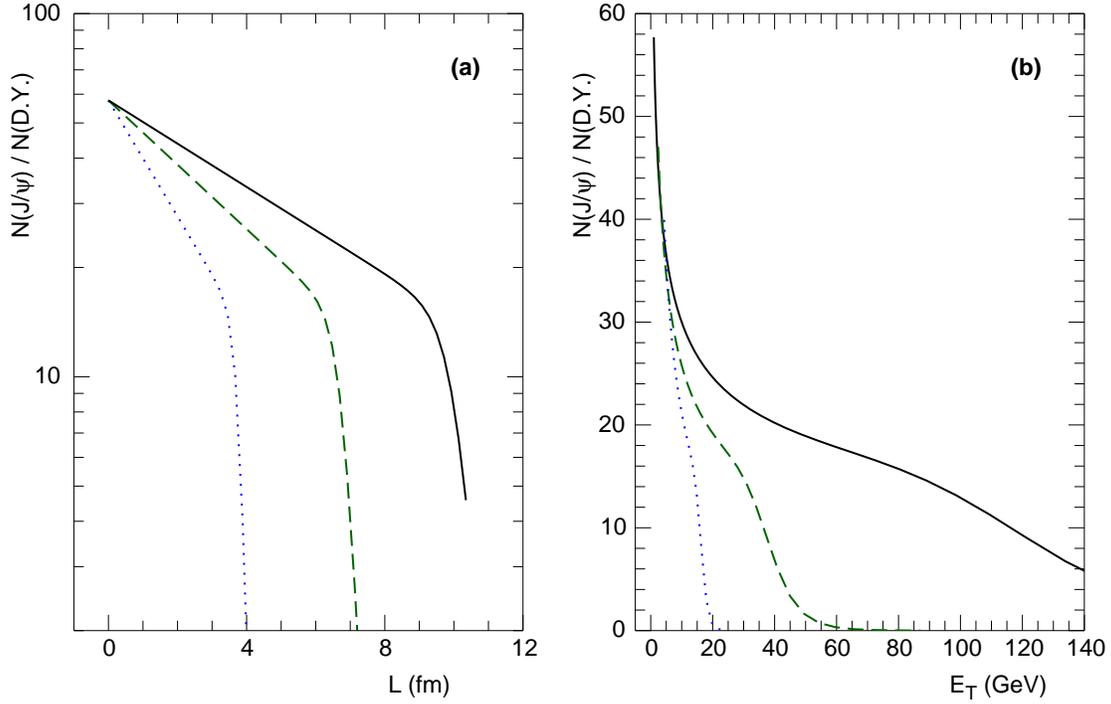}}
\end{center}
\caption[comparison with data]{The ratio of \Jpsi\ to Drell-Yan events
as predicted by eq.~\eqref{eq:RL}, predicted at RHIC energy (dashed
lines) and at LHC energy (dotted lines), compared with SPS energy
(solid lines).}\label{fig:prediction}
\end{figure}

Finally, we present in Figure~\ref{fig:prediction}
our predictions for  RHIC and LHC energies,
compared with the curve for SPS. Figure~\ref{fig:prediction}a displays
eq.~\eqref{eq:Rfull} with the expected change in $\rho_s$ determined by the
change in the number of strings given in \cite{Pajares:perc}, 
and with the other parameters ($\sigma, r_s$) held constant, 
In Figure~\ref{fig:prediction}b we
transformed the curve from the $L$ variable to $E_T$ with the
conservative assumption that the proportionality between $E_T$ and the
number $\nu$ of elementary collision is independent of $\sqrt{s}$.
This implies that
\begin{equation}
  E_T(L;\sqrt{s}) = \frac{\rho_s(\sqrt{s})}{\rho_s(\sqrt{s_0})} 
			E_T(L;\sqrt{s_0})
\end{equation}
and we use eq.~\eqref{eq:L-Et} for $\sqrt{s_0} = 19.4$ GeV. We see
that, as $\sqrt{s}$ grows,
 both the absorptive part gets larger and the change of curvature
denoting percolation sets in at a much smaller value of $E_T$.

\section*{Acknowledgements}
This work was partially supported by project 
PRAXIS/PCEX/P/FIS/124/96.\linebreak
R.U. gratefully acknowledges the financial support of the
Fundação Ciência e Tecnologia via the ``Sub-Programa Ci\^encia 
e Tecnologia do $2^o$ Quadro Comunit\'ario de Apoio.''

\section*{References}


\begin{thebibliography}{15}

\bibitem{components}		
J. Dias de Deus, C. Pajares and C.A. Salgado,   
 Phys.\ Lett.\  B\,407 (1997) 335; 
 Phys.\ Lett.\  B\,408 (1997) 417.


\bibitem{rare}		
J. Dias de Deus, C. Pajares and C.A. Salgado,   
 Phys.\ Lett.\  B\,409 (1997) 474.

\bibitem{nuclear_absorption}		
B. Blakenbecker, A. Capella, J. Tran Thanh Van, C. Pajares and A.V. Ramallo,    
 Phys.\ Lett.\  B\,107 (1981) 106.

\bibitem{QGP}		
T. Matsui and H. Satz,   
 Phys.\ Lett.\  B\,178 (1986) 416.

\bibitem{NA50:Jpsi}		
M.C. Abreu et al. (NA50 Collaboration),    
 Phys.\ Lett.\  B\,450 (1999) 456.

\bibitem{Jpsi:absorption}		
A. Capella, C. Merino, J. Tran Thanh Van, C. Pajares and A.V. Ramallo,    
{Phys.\ Lett.\ }{B}{43} (1990) {144}; 
S. Gavin and R. Vogt,   
{Phys.\ Rev.\ Lett.\ }{}{78} (1997) {1006}; 
A. Capella, A. Kaidalov, A. Kouider Akil, C. Gerschel,   
 Phys.\ Lett.\  B\,393 (1997) 431.

\bibitem{DPM:1}		
A. Capella, U.P. Sukhatme, C.I. Tan and J. Tran Thanh Van,   
 Phys.\ Rep.\  236 (1994) 225.

\bibitem{ccstrings}  
J. Geiss, C. Greiner, E.L. Bratkovskaya, W. Cassing and U. Mosel,
 Phys.\ Lett.\ B\,447 (1999) 31. 

\bibitem{perc:previous}		
G. Baym, 
{Physica }{A}{96} (1979) {131};  
T. Celik, F. Karsch and H. Satz, 
 Phys.\ Lett.\  B\,97 (1980) 128.

\bibitem{Pajares:perc}		
N. Armesto, M.A. Braun, E.G. Ferreiro and C. Pajares,    
 Phys.\ Rev.\ Lett.\  77 (1996) 3736.

\bibitem{perc:98}		
A. Rodrigues, R. Ugoccioni and J. Dias de Deus,   
 Phys.\ Lett.\  B\,458 (1999) 402.


\bibitem{NA50:pp}		
M.C. Abreu et al. (NA50 Collaboration),    
 Phys.\ Lett.\  B\,438 (1998) 35.

\bibitem{NA50:L}		
M.C. Abreu et al. (NA50 Collaboration),    
 Phys.\ Lett.\  B\,410 (1997) 337.

\bibitem{ccbarlength}		
C. Gerschel and J. H\"ufner,   
 Z.\ Phys.\  C\,56 (1992) 171.

\bibitem{DiasDeus:Jpsi}		
J. Dias de Deus and J. Seixas,    
 Phys.\ Lett.\  B\,430 (1998) 363.

\bibitem{NA50:QM99}     
C. Cical\`o (NA50 Collaboration), talk presented at ``Quark Matter 99''
  (Torino, Italy, May 10--15, 1999).

\end{thebibliography}
\end{document}